# High Sensitivity φ-OTDR over Long Distance with Polarization Multiplexed Codes

Christian Dorize, Elie Awwad, *Member, IEEE*, and Jérémie Renaudier, *Member, IEEE*

*Abstract*—Newly introduced polarization diversity probing codes are suggested to enhance the sensitivity and bandwidth performance of differential phase-sensitive distributed OTDR systems. This was recently demonstrated by means of short-length specialized fibers with a backscattering induced by equally spaced Fiber Bragg Gratings inserted in the fiber itself. The work summarized in this letter aims to extend the polarization-diversity probing technique to widely spread standard single mode fibers (SSMF) used for telecommunications, by solely exploiting the Rayleigh backscattering. Conditions to achieve perfect phase estimation along such fibers are first detailed. An experimental validation highlights the ability to detect and localize, over 25km of SSMF, multiple low-energy perturbations within a 475Hz-bandwidth.

## I. Introduction

There is a growing interest for in-depth monitoring of mechanical events in many fields, mainly driven by security and automation applications. Thanks to the huge amount of fibers already deployed for telecom transmission especially in urban regions, Distributed Acoustic Sensing (DAS) by means of these telecom fibers is very appealing to avoid installing dedicated fibers or discrete components such as electro-mechanical sensors all along the streets, railways or in buildings. In addition, DAS technology has evolved from solely relying on optical intensity to newly introduced coherent phase-sensitive scheme that uses a coherent mixer [1, 2], enabling spatially distributed sensing for a wider range of mechanical events thanks to enhanced linearity and sensitivity, and an extended bandwidth. To further enhance the performance of coherent phase-sensitive DAS, we recently proposed an interrogator that jointly probes two orthogonal polarization states by means of mutually orthogonal complementary Golay pairs [3] and detects the backscattered light using a dual-polarization coherent mixer [4]. The performance of this new interrogator was measured over a dedicated short-length fiber sensor array made of equally distributed 0.1% reflecting Fiber Bragg Gratings (FBGs) [4, 5]. This letter aims to extend the scope of the interrogator to standard single mode fibers (SSMFs) by detecting weak and randomly distributed Rayleigh backscattering. We demonstrate the ability to detect and localize mechanical vibrations all over a 26km SSMF within a bandwidth of 475Hz, with a distance dependent sensitivity of $15nm_{pp}$ at 1km and $40nm_{pp}$ at 25km. Such performance figures allow for high-precision monitoring of vibration events impacting installed fiber infrastructure.

## II. Polarization diversity with Rayleigh backscattering

In most recent DAS systems, the interrogator is made of light pulses or sweep signals [6, 7] transmitted periodically on a single polarization axis. In these schemes, the backscattered signal is subject to polarization fading effects with sub-optimal performance in sensitivity, maximum reach or bandwidth for the detection of mechanical events present along the sensing fiber. Polarization fading was tackled in recent works using a polarization-diversity coherent receiver to detect the information on the state of polarization of the backscattered signal by projecting it on the two orthogonal polarization states of the receiver as in [8]. We introduced in [4] a polarization-diversity interrogation technique that simultaneously probes two polarization axes by means of two mutually orthogonal binary complementary pairs, along with a dual-polarization coherent receiver to capture the backscattered signals. This method gives access to the dual-pass Jones matrices up to a given location in the fiber while also mitigating polarization fading. The phase information is extracted from the estimated Jones matrices as explained later in this section. The conditions for achieving a perfect fiber impulse response estimation were studied in [4]. To validate our new interrogation scheme, two scenarios were considered. In a former work [5], a dedicated fiber was used in which fiber Bragg gratings (FBGs), enhancing the intensity of the backscattered light, were regularly distributed. In this work, we considered randomly distributed reflectors which covers Rayleigh backscattering randomly spread over SSMFs as those deployed in metropolitan optical networks and submarine long-haul transmissions. The polarization-multiplexed probing signals that are used in this paper are reminded below.

Two mutually orthogonal complementary binary pairs {$G_{a1}$, $G_{b1}$} and {$G_{a2}$, $G_{b2}$} are used to probe the fiber. The first pair modulates a given polarization state of the optical field (linear horizontal state for instance) while the second pair modulates a state that is orthogonal to the first (linear vertical). The two pairs

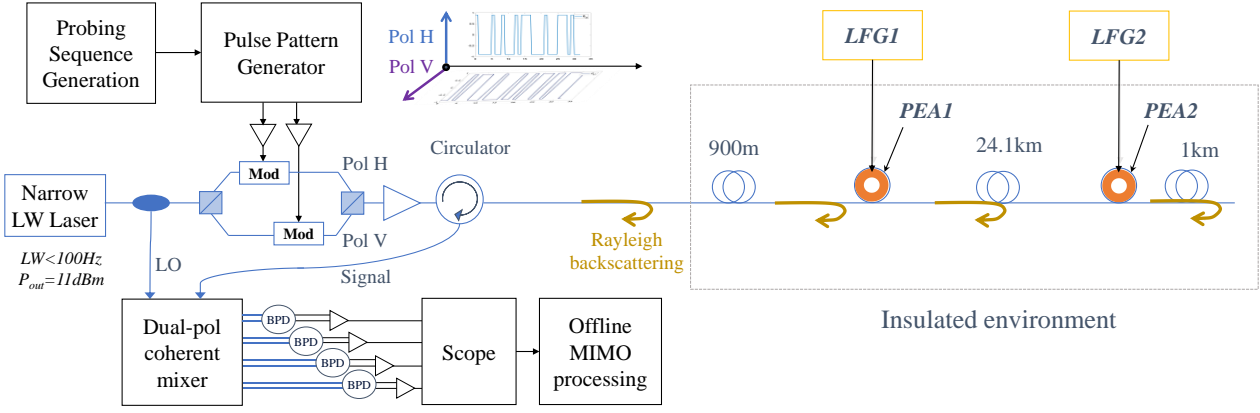

Fig.1. Experimental setup of the interrogation scheme (BPD: Balanced Photodiode, LFG: Low Frequency Generator, LW: Linewidth, MIMO: Multiple-Input Multiple-Output, Mod: Modulator, PEA: Piezoelectric Actuator)

simultaneously modulate the two polarization states through a polarization-division-multiplexed binary phase-shift-keying (BPSK) mapping. A basic example of mutually orthogonal complementary pairs [9] is given by $G_{a1}=[1,-1,-1,-1]$, $G_{b1}=[-1,1,-1,-1]$, $G_{a2}=[-1,-1,1,-1]$, $G_{b2}=[-1,-1,-1,1]$. Longer sequences can be obtained recursively from the above seed to get the desirable probing length [3]. The duration of the probing code is defined as $T_{code}=2 \cdot (4 \cdot 2^K)/F_{symb}$, where $K$ is an integer standing for the number of recursions and $F_{symb}$ is the fixed symbol rate. The sensing fiber is continuously probed with a period $T_{code}$ yielding a mechanical bandwidth $BW=1/(2 \cdot T_{code})$. The native spatial resolution $S_r$ is equal to $c_f/(2F_{symb})$, where $c_f=c/n$ is the velocity of light in the SSMF core with refractive index $n \approx 1.45$ and $c$ being the speed of light in vacuum.

At the receiver, the backscattered signal from the $i^{th}$ segment and at a $j^{th}$ time index is given by $\boldsymbol{E_r} = \boldsymbol{H}_{i,j}\boldsymbol{E_t}$ with $\boldsymbol{E_t} = \begin{pmatrix} E_{tx} \\ E_{ty} \end{pmatrix}$, $\boldsymbol{E_r} = \begin{pmatrix} E_{rx} \\ E_{ry} \end{pmatrix}$ being the transmitted and received optical fields respectively onto each of two orthogonal polarization axes $x$ and $y$, while $\boldsymbol{H}_{i,j}$ is the dual-pass impulse response up to the $i^{th}$ fiber segment at time index $j$ represented by a 2x2 Jones matrix. We neglect in this representation all dispersion effects (chromatic and polarization mode dispersion) given that the covered fiber distances are shorter than 100km and the used symbol rates are low compared to the ones in standard optical fiber telecommunications. A multiple-input multiple-output (MIMO) processing consisting of four correlations between each of the two received optical fields $\{E_{rx}, E_{ry}\}$ and each of the two transmitted codes $\{E_{tx}, E_{ty}\}$ is performed to get the $\boldsymbol{H}_{i,j}$ estimations periodically at each $T_{code}$. Even though an information on the evolution of the state of polarization (SOP) can be derived from these matrices, we only concentrate here on the optical phase, extracted from the matrix $\boldsymbol{H}_{i,j}$ as $\varphi_{i,j}=0.5 \, angle(det(\boldsymbol{H}_{i,j}))$ where $det()$ stands for the determinant of $\boldsymbol{H}_{i,j}$. Having computed the absolute phases up to the $i^{th}$ segment, the phase evolution per segment is obtained from the differential phases with setting the phase from the first reflector as a reference.

The conditions for perfect estimation of $\boldsymbol{H}_{i,j}$ from $\boldsymbol{E_r}$ when the sensing fiber is probed with polarization-division-multiplexed PDM-BPSK codes [4] is $T_{code} > 4 \cdot T_{ir}$ where $T_{ir}=2L/c_f$ stands for the time spreading of the channel response for a fiber of length $L$. Therefore, the symbol rate and the code length must be chosen accordingly to come up with a $T_{code}$ value that complies with the above condition. Notice that selecting a $T_{code}$ close to the lower limit $4 \cdot T_{ir}$ allows for a maximization of the mechanical bandwidth $BW$ of the sensing system. Conversely, a longer $T_{code}$ results in an enhanced signal-to-noise ratio after the correlation process at the receiver, thus improving the sensitivity, i.e. the minimal detectable fibre mechanical extension magnitude. In practice, a supplementary condition over $T_{code}$ is enforced. Its value should remain shorter than the coherence time $T_{coh}$ of the laser source to ensure a reliable extraction of the phase information [3]. The coherence time of a laser source is mainly determined by its spectral linewidth. Lasers with a linewidth narrower than 1kHz are required to cover a round-trip distance around 100km. In summary, with Rayleigh backscattering, the probing code duration determines a trade-off between the bandwidth and the sensitivity and must be selected within a restricted range such as $4 \cdot T_{ir} < T_{code} < T_{coh}$, the lower limit being imposed by the defined polarization-multiplexed BPSK-coded Golay pairs.

## III. EXPERIMENTAL SETUP

The interrogator unit and experimental setup are shown in Fig.1. A narrow linewidth (LW<100Hz) laser source emitting 11 dBm at $\lambda$=1536.6nm is used to generate the interrogation signal at the transmitter side and serves as a local oscillator at the receiver side. The two BPSK complementary probing sequences modulate the optical signal through a dual-polarization Mach-Zehnder modulator (MZM) at $F_{symb}$=125MBaud. The polarization-multiplexed optical signal is sent through an optical amplifier and a circulator into three interconnected SSMF fiber spools of lengths 0.9, 24.1 and 1 km respectively, leading to an overall fiber length of 26 km. Two independent perturbations are inserted on two fiber segments between these spools through cylindrical piezoelectric actuators having an external diameter of 5cm. 55cm (resp. 133cm) of fiber loops are coiled around the first (resp. second) actuator. The backscattered Rayleigh signal is sent to a dual-polarization coherent mixer consisting of two 90° optical hybrids. The in-phase and in-quadrature beating signals projected over two orthogonal polarization states are then captured by balanced photodiodes. The resulting photocurrents are amplified through transimpedance amplifiers (TIAs) and captured on a 12bit-scope with a sampling rate of 250MSample/s. Acquired data is then transferred on a PC for offline processing which consists in correlating the received signals with the probing sequences to extract Jones matrices from which a space-time map of the optical phase is obtained. Fig.2 shows the Rayleigh backscatter intensity as a function of the distance for a 1s long measurement time, highlighting the standard 0.2dB/km loss in the SSMF spool. We also notice the slight local losses at the connectors between these spools, mostly for the last connection.

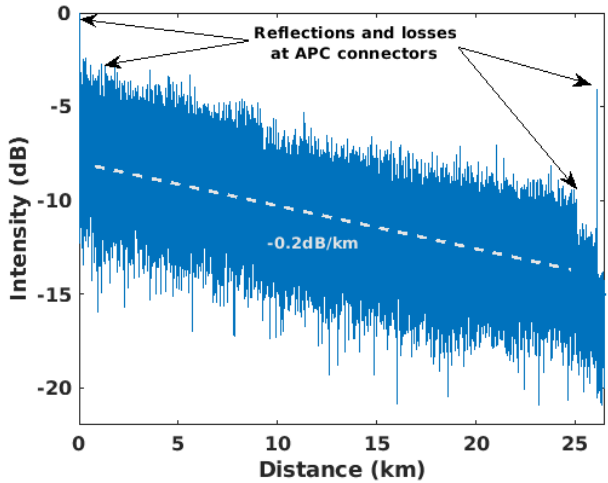

Fig. 2. Intensity of Rayleigh backscattered signal along the 26km SSMF

## IV. EXPERIMENTAL RESULTS

We demonstrate the distributed sensing ability over a long distance by jointly perturbating two fiber segments. 300Hz and 180Hz electrical sine waves of same magnitude are injected into the two actuators located at 900m and 25km respectively. The 26km optical line is continuously probed with 1.05ms-long PDM-BPSK codes corresponding to $2^{17}$-symbol sequences at 125MBaud, providing a 475Hz mechanical bandwidth. Prior to introducing these mechanical perturbations, we first estimate the phase stability by measuring the standard deviation (StDv) in time of the differential phases calculated from a subset of Jones matrices separated by 100m on average along the 26km fiber. Fig.3(a) displays in dash-dot lines 5 successive 1s-long static measurements separated by a 1-minute interval each. The phase StDv slowly increases with the distance, from 0.02 to 0.2 rad. Superimposed in bold blue is the StDv obtained in dynamic mode with the two generated perturbations. A zoom is displayed at each actuator position highlighting the joint detection and localization of the two independent mechanical events. Notice that the standard deviation ratio of the two averaged peaks (0.22 and 0.52 rad) complies with the ratio between the lengths of the two fiber sections where the vibrations are applied (55 and 133cm respectively).

Fig.3(b) shows the differential phases as a function of time over a subsection of the fiber centered at the first actuator position (900m): the detected 300Hz sine wave perturbation shows a peak to peak excursion of 0.57rad whereas the differential phase at the two displayed surrounding locations is very stable in time, highlighting the lack of crosstalk. Similarly, Fig.3(c) displays the phase versus time around the position of the next perturbation at 25km. The applied 180Hz sinusoidal strain is here again well detected, with a 1.25rad$_{pp}$ excursion. However, we notice a much higher noise level at the two displayed surrounding locations when comparing to the static phases of Fig.3(b), which reflects the SNR loss induced by the increased fiber distance.

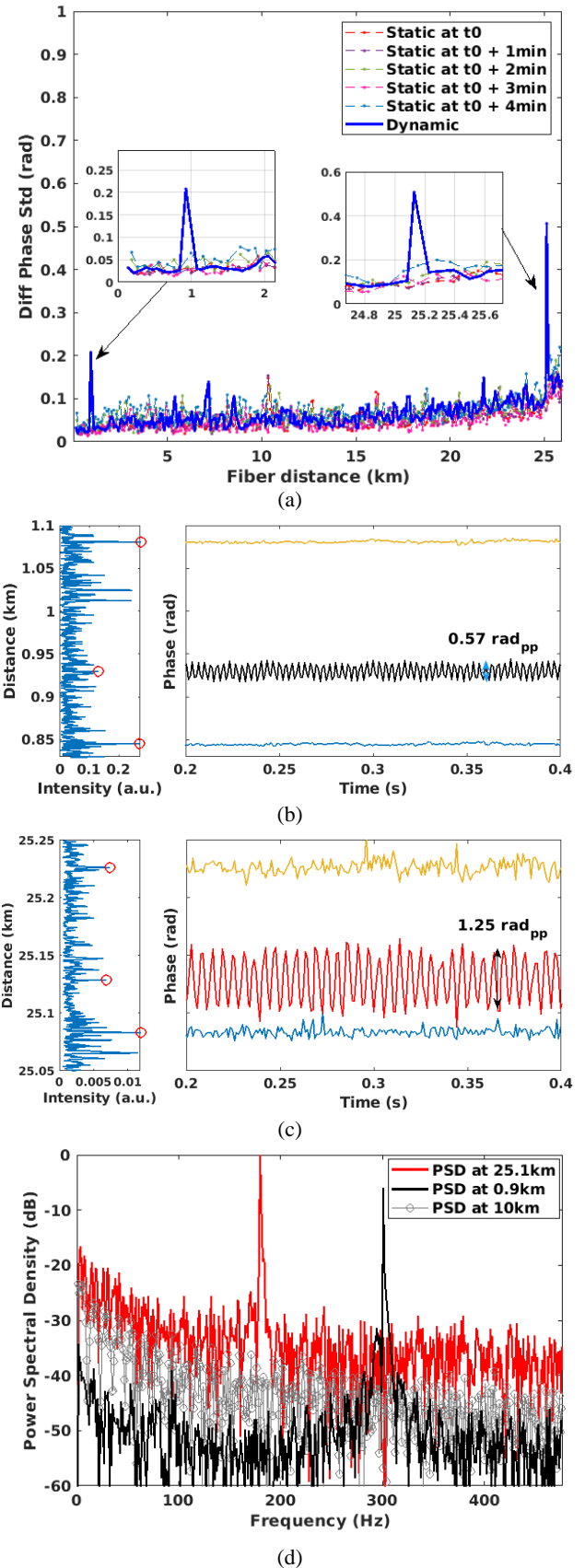

Fig.3. (a) Differential phase standard deviations over 26 km. (b&c) Differential phase as a function of time around each of the two piezoelectric locations. (d) Power spectral densities at the piezoelectric locations and at an unperturbed intermediate segment at 10km.

The power spectral density (PSD) of the differential phase computed at the location of each actuator is shown in Fig.3(d), with an SNR value over 50 (resp. 30) dB at the first (resp. second) piezoelectric actuator. We also computed the PSD at several intermediate fiber locations (before, after and between the two perturbations) to confirm the lack of emerging frequency component there (no spatial crosstalk). We display in Fig.3(d) the PSD of the phase at 10km; the obtained noise density illustrates the spectral distribution of the self-noise of the setup.

Finally, we study the sensitivity at each of the two locations where the perturbation is applied using the same actuator. A 100Hz sinewave is injected in the actuator with a variable voltage from 0.05 to 18.8V$_{pp}$, leading to fiber extensions in a range of 1 to 425nm$_{pp}$. This latter maximum extension is imposed by our current setup (highest voltage at the output of our low-frequency generators), hence preventing us from fully exploring the dynamic range of the system. This limit can be surmounted in a future study to measure the dynamic range by winding more fiber spools around the piezo or using an electrical amplifier. Fig.4 displays the peak-to-peak phase magnitude of the captured 100Hz sinewave as a function of the fiber extension for each of the two considered locations, measured over a 0.2s window. At 900m, we notice a linear increase from a 15nm$_{pp}$ magnitude perturbation; below the observed 0.09rad$_{pp}$ threshold, which translates into a 0.03rad StDv noise floor, the perturbing sinewave is lost in the noise and cannot be detected without further time averaging. At 25km, the phase magnitude matches as expected the one measured at 900m but only beyond a perturbation of 40nm$_{pp}$ magnitude, whereas the noise floor StDv is evaluated to 0.14rad. This higher value is here again a consequence of SNR loss induced by the 0.2dB/km attenuation in the fiber; the noise floor increase is also visible in the standard deviation evolution of the phase displayed in Fig.3(a). Beyond their respective noise floor, notice that the two measurements match the phase variation versus the fiber extension $dL$ reference curve given by $d\varphi = dL \cdot 4\pi n\xi/\lambda$, where $\xi=0.78$ is the photo-elastic scaling factor for a longitudinal fiber strain [10]. Though it could not be fully characterized due to our current setup 18.8V$_{pp}$ maximal excitation voltage, the power dynamic range spreads at least over 33dB and 20dB at 900m and 25km respectively. These values show that reasonably high dynamic range mechanical events may be captured from the optical phase up to long distances in SSMFs without compression or clipping effects.

## V. CONCLUSION

We studied the ability to detect and localize mechanical events over a long fiber distance solely from Rayleigh backscattering thanks to recently introduced PDM-BPSK codes. It was shown that two independent mechanical events, separated here by 24.1km, can be jointly detected and localized within a mechanical bandwidth of 475Hz. The sensitivity at fiber distances of 900m and 25km was estimated to 15nm$_{pp}$ and 40nm$_{pp}$ respectively. The achieved sensitivity versus bandwidth performance over long distances paves the way for a wide range of new surveillance and monitoring applications using already deployed telecom fibers.

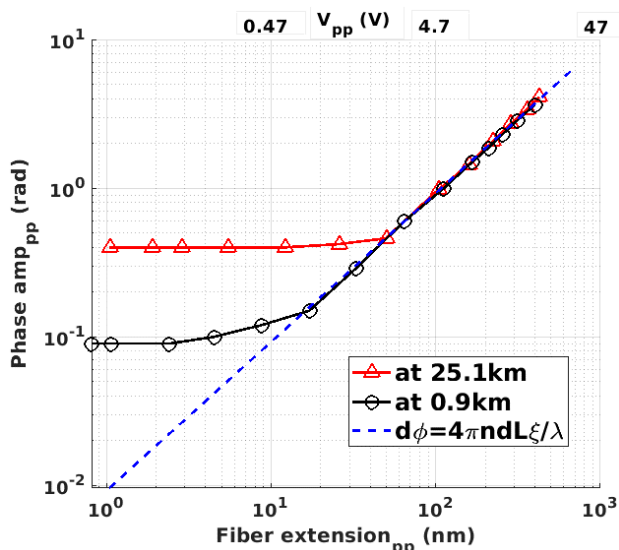

Fig. 4. Peak-to-peak phase magnitude as a function of the applied peak-to-peak fiber extension at the locations of the two piezoelectric actuators.